\documentclass[prb,reprint,twocolumn,showpacs]{revtex4}
\usepackage{amsmath,amssymb,graphicx,xcolor,units,hyperref}

\newcommand{\dd}[1]{\mathrm{d}#1\,}
\newcommand{\avg}[1]{\langle{#1}\rangle}
\renewcommand{\Re}{\mathop{\mathrm{Re}}}

\newcommand{\Tr}{\mathop{\mathrm{Tr}}}

\newcommand{\tr}{\mathop{\mathrm{tr}}}

\newcommand\varpm{\mathbin{\vcenter{\hbox{%
  \oalign{\hfil$\scriptstyle+$\hfil\cr
          \noalign{\kern-.3ex}
          $\scriptscriptstyle({-})$\cr}%
}}}}
\newcommand\varmp{\mathbin{\vcenter{\hbox{%
  \oalign{$\scriptstyle({+})$\cr
          \noalign{\kern-.3ex}
          \hfil$\scriptscriptstyle-$\hfil\cr}%
}}}}

\DeclareMathOperator{\arctanh}{arctanh}

\begin{document}

\title{Thermal transport through ac-driven transparent Josephson weak links}

\author{P. Virtanen}
\affiliation{O.V. Lounasmaa Laboratory, Aalto University,
  P.O. Box 15100,FI-00076 AALTO, Finland}

\author{F. Giazotto}
\affiliation{NEST, Istituto Nanoscienze-CNR and Scuola Normale Superiore, I-56127 Pisa, Italy}

\date{\today}


\begin{abstract}
  We discuss how phase coherence manifests in the heat transport
  through superconducting single and multichannel Josephson junctions in time
  dependent situations. We consider the heat current driven through
  the junction by a temperature difference in dc voltage and ac phase
  biased situations.  At low bias, the electromagnetic driving mainly
  modifies the form of the coherent resonance that transports a large
  part of the heat current, rather than simply dissipating energy in
  the junction.  We find a description for the heat current in terms
  of quasiparticle $n$-photon absorption and emission rates, and
  discuss analytical and numerical results concerning them. In
  addition to the ensemble average heat transport, we describe also
  its fluctuations.
\end{abstract}

\maketitle

\section{Introduction}

Time-dependent effects in superconducting junctions have been under
extensive study since the discovery of the Josephson effect. As far as
charge transport is considered, the physics is well understood both
theoretically and experimentally and has been studied in detail ---
both for tunnel junctions \cite{werthamer1966-nso} and for finite
junction
transparencies. \cite{averin1995-aje,zaitsev1998-toa,zazunov2003-alq,gorelik1995-mis,kos2013-fda,bergeret2010-tom,bergeret2011-saa}
In contrast, time-dependent heat transport through Josephson junctions
has attracted interest only quite recently, \cite{golubev2013-htt} in
the wake of theoretical
\cite{maki1965,guttman1997-pdt,guttman1998-ieh,zhao2003-phase,zhao2004-htt,golubev2013-htt,virtanen2014-fhc}
and experimental \cite{giazotto2012-jhi,martinez2014-qdt} advances in
understanding of how the heat transport is modulated by phase
differences of the superconducting order parameter.

\begin{figure}
  \includegraphics{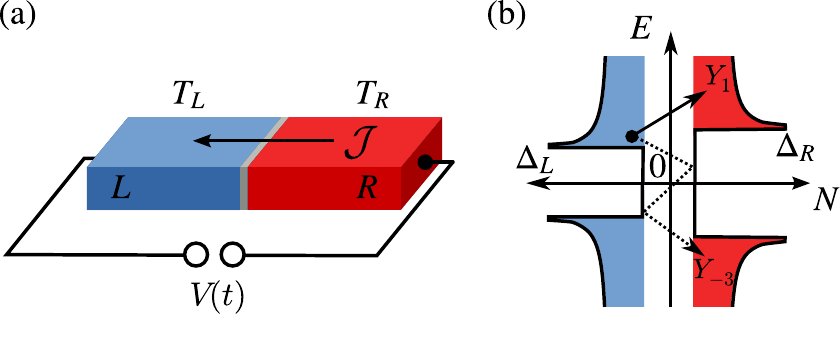}
  \caption{\label{fig:heatdriven-setup} 
    (a) 
    Heat current ${\cal J}$
    flows through a Josephson junction between identical 
    superconductors $R$ and
    $L$, driven by a temperature difference $T_R>T_L$.  The
    simultaneously applied time-dependent voltage $V(t)$ not only
    dissipates power in the junction, but also modulates the phase
    coherent effects related to quasiparticle transport, thereby
    changing ${\cal J}$ even when the dissipated power is
    negligible compared to ${\cal J}$.  
    (b) 
    Heat is carried by quasiparticles traversing the
    junction.  They can absorb or emit energy by interacting with the
    electromagnetic field in the junction.  For monochromatic ac bias
    $V(t)\propto\sin(\omega{}t)$, energy absorbed is an integer $k$
    multiple of the frequency. Heat transport is fully described by the
    total rates $Y_{k}(E)$ of the corresponding processes.  
    $\Delta_L$ and $\Delta_R$ denote the temperature-dependent energy gaps
    of the two superconductors.
  }
\end{figure}

In a stationary situation, electronic heat current between two
superconductors connected by a weak link is mainly mediated by the
transport of above-gap quasiparticles.  The heat transport is
determined by their energy dependent transmission rates $Y(E)$,
\cite{maki1965,guttman1997-pdt,guttman1998-ieh,zhao2003-phase,zhao2004-htt}
which in the simplest picture can be found by solving the
Bogoliubov--de Gennes scattering problem associated with the weak
link. \cite{virtanen2014-fhc,beenakker1991-ulc} Due to the presence of
phase-coherent Andreev reflections, the transport depends on the phase
difference between the superconducting order parameters of the two
sides, and moreover exhibits transmission resonances which are of
importance especially in low-transparency
junctions. \cite{maki1965,guttman1997-pdt,guttman1998-ieh,zhao2003-phase,zhao2004-htt,golubev2013-htt,virtanen2014-fhc}

Presence of an electromagnetic field in the weak link enables
additional possibilities for quasiparticle transfer, as energy can be
absorbed from or emitted to the field [see
Fig.~\ref{fig:heatdriven-setup}]. Transport is no longer elastic, and
processes such as photon assisted tunneling \cite{tucker1985-qda} and
multiple Andreev reflections \cite{tinkham1996-its,octavio1983-ses}
appear.  The energy absorption physics is also related to the
quasiparticle dynamics of the Andreev bound states (ABS) residing in
the junction. This physics has been studied by a number of
recent experiments using quantum point contacts with a small number of
transport channels,
\cite{zgirski2011-elq,bretheau2013-eap,bretheau2013-ssa} diffusive
metal junctions, \cite{chiodi2011-pdo,fuechsle2009-eom} and phase dependent
absorption has been studied in tunnel junctions \cite{pop2014-cse}.

To describe the effects of time-dependent drive on the heat current
carried by quasiparticles, the energy dependent transmission rates
$Y(E)$ of the stationary state need to be generalized to depend on
both the initial $E'$ and final $E$ quasiparticle energies $Y(E,E')$.
If the junction is driven with a periodic signal at
frequency $\omega_0$ (or dc voltage $eV=\hbar\omega_0$), the final
energy is $E=E'-k\omega_0$, and the results can be described in terms
of a set of rates $\{Y_k(E)\}$ related to emission/absorption of $k$
photons to the field, as illustrated in
Fig.~\ref{fig:heatdriven-setup}(b).

Heat transport and its fluctuations in the presence of time-dependent
driving have been recently discussed in terms of Floquet scattering
matrices in Refs.~\onlinecite{moskalets2014-fsm,battista2014-eap} for
normal-state systems. In this description, the rates $\{Y_k(E)\}$ can
be directly related to the Floquet scattering matrices. Although these
works consider normal-state systems, the formal description of
electronic heat current in time-dependent superconducting junctions
turns out to be similar.  This should be contrasted to charge current:
the superconducting condensate transports charge without requiring
quasiparticle transfer, which implies that transport of charge and
heat are not as closely connected in superconductors as in
normal-state systems.

In this work, we study theoretically the heat transport in
superconducting junctions of general transparency driven by classical
fields, as shown in Fig.~\ref{fig:heatdriven-setup}.  We derive
results for the rates $\{Y_k(E)\}$ and for the heat transport and its
fluctuations using quasiclassical Green function theory. Previously, a
similar approach has been used in Ref.~\onlinecite{kopnin2008-ipa} for
studying time-dependent heat transport in NIS junctions, and it has
been extensively applied for examining the charge transport (see
e.g. Refs.~\onlinecite{zaitsev1998-toa,cuevas2004-dti,bergeret2010-tom,bergeret2011-saa}). For
tunnel junctions, the approach here is equivalent to that of
Ref.~\onlinecite{golubev2013-htt}, but it can also describe the
time-dependent heat transport in junctions with a finite transparency
away from the tunneling limit.

We consider situations with dc and ac bias, and discuss how the
electromagnetic driving modulates the temperature-driven heat
current. In particular, for weak driving, we find the effect is mainly
dominated by a modulation and phase-averaging of the coherent
resonances appearing in the rates $\{Y_k(E)\}$. We find analytical
descriptions in certain limiting cases, and discuss representative
features in terms of numerically exact results.

This manuscript is structured as follows.  Section \ref{sec:model}
introduces the theoretical and numerical approaches used.  Section
\ref{sec:voltage-bias} discusses the effect of a dc voltage bias on
the heat transport, and \ref{sec:ac-bias} concentrates on the ac bias
case. Section \ref{sec:discussion} summarizes the main points.  We
postpone details of the derivation of the analytical results to a
number of appendixes, referring to the salient points in the main
text.

\section{Model}
\label{sec:model}

We consider the situation depicted in Fig.~\ref{fig:heatdriven-setup}:
heat transport between two identical superconducting terminals that
are connected by a generic contact described by the transmission
eigenvalues $\{\tau_j\}$ of the normal-state scattering matrix of the
junction.  The terminals are kept at different constant temperatures,
$T_L$ and $T_R$, and a time dependent voltage or phase bias is applied
between them. Moreover, they are assumed to have sufficient impurity
concentration to lie in the dirty limit.  To study the transport in
this system, we make use of the Keldysh-Nambu Green function
formulation for transport in superconducting structures,
\cite{larkin1986-ns,belzig1999-qgf,usadel1970-gde} and the
quasiclassical boundary condition description of a weak link between
bulk superconductors in the diffusive limit.
\cite{zaitsev84,zaitsev1998-toa,nazarov99}

At equilibrium, electrons inside a superconducting terminal at
temperature $T$ with superconducting gap $\Delta$ are described by the
quasiclassical equilibrium Green function $\check{g}_0(E)$,
\begin{align}
  \label{eq:g-equilibrium}
  \check{g}_0 
  &= 
  \begin{pmatrix}
    \hat{g}^R_0 & (\hat{g}^R_0-\hat{g}^A_0)\tanh\frac{E}{2T} \\ 0 & \hat{g}^A_0
  \end{pmatrix}
  \,,
  \\
  \hat{g}^R 
  &= 
  \begin{pmatrix}
    \frac{E}{\sqrt{(E+i\eta)^2 - |\Delta|^2}} & \frac{\Delta}{\sqrt{(E+i\eta)^2 - |\Delta|^2}} \\     
    -\frac{\Delta^*}{\sqrt{(E+i\eta)^2 - |\Delta|^2}} & -\frac{E}{\sqrt{(E+i\eta)^2 - |\Delta|^2}}
  \end{pmatrix}
  \,,
\end{align}
with $\eta\to0^+$ and $\hat{g}^A=-\hat{\tau}_3\hat{g}^R\hat{\tau}_3$, where
$\hat{\tau}_3$ is the third spin matrix in the Nambu space.  We assume
the energy gap $\Delta(T)$ depends on the temperature according to the
BCS gap relation. \cite{tinkham1996-its} The Green function describing
a superconductor at a nonzero electric potential $V$ is found by a
gauge transform
\begin{align}
  \label{eq:biased-terminal}
  \check{g}(t,t') =
  e^{i\hat{\tau}_3\phi(t)}\check{g}_0(t-t')e^{-i\hat{\tau}_3\phi(t')}
  \,,
\end{align}
where $\phi(t) = \frac{e}{\hbar} \int^t\dd{t'} V(t')$ is the
electromagnetic phase. \cite{zaitsev1998-toa,kopnin2001-ton} Here,
$\check{g}_0(t-t')=\int_{-\infty}^\infty\frac{\dd{E}}{2\pi}e^{-iE(t-t')}\check{g}_0(E)$.
We consider the phase as a classical field, neglecting its quantum
fluctuations, which simplifies its Keldysh structure.

For cases where the frequencies in the drive $\phi$ and the dc bias
voltage are commensurate, with a base harmonic $\omega_0$, we can use
a (Floquet) matrix representation \cite{cuevas2006-pea},
\begin{align}
  \label{eq:matrixrepr}
  \check{g}(t,t') =
  \sum_{k,m=-\infty}^\infty
  \int_0^{\omega_0}\frac{\dd{\epsilon}}{2\pi}
  e^{-i(\epsilon+m\omega_0)(t-t')-ik\omega_0t}\check{g}_{m+k,m}(\epsilon)
  \,.
\end{align}
In this representation, time convolution
$\int_{-\infty}^\infty\dd{t_1}A(t,t_1)B(t_1,t)$ is equivalent to a
matrix product, $\sum_k A_{nk}(\epsilon_0) B_{km}(\epsilon_0)$.
Below, products between two-time quantities
with time or energy arguments omitted imply time convolutions. We also
denote $\phi(t,t')\equiv\delta(t-t')\phi(t)$, so that
Eq.~\eqref{eq:biased-terminal} is written as
$\check{g}=e^{i\hat{\tau}_3\phi}\check{g}_0e^{-i\hat{\tau}_3\phi}$.

The charge and heat currents flowing between two connected
superconductors can be found from boundary conditions applicable to
quasiclassical Green functions. \cite{zaitsev1998-toa,nazarov99}
Transport properties of the junction are described by a matrix
current \cite{nazarov99}
\begin{align}
  \check{I}
  =
  [\check{g}_L, \check{g}_R]_- \check{Z}
  \,,\;
  \check{Z}
  =
  \frac{e^2}{h}
  \sum_n 
  \frac{
    \tau_n
  }{
    1 - \frac{\tau_n}{2}
    +
    \frac{\tau_n}{4}[\check{g}_L, \check{g}_R]_+
  }
  \,,
\end{align}
where $\check{g}_L$ and $\check{g}_R$ are the Green functions inside
the terminals at the two sides of the junctions. In particular, the
charge current, and the heat current entering terminals $L$ and $R$
read (cf. also Ref.~\onlinecite{kopnin2008-ipa}),
\begin{align}
  I(t)
  &=
  \frac{1}{8}
  \tr\check{\sigma}_1\hat{\tau}_3\check{I}(t,t)
  \,,
  \\
  \label{eq:QLdef}
  \dot{Q}_L(t)
  &=
  \frac{1}{16}
  \tr\check{\sigma}_1[
  (\epsilon+\mu\hat{\tau}_3)\check{I} + \check{I}(\epsilon+\mu\hat{\tau}_3)
  ]
  (t,t)
  \,,
  \\
  \label{eq:QRdef}
  \dot{Q}_R(t)
  &=
  -\frac{1}{16}
  \tr\check{\sigma}_1[
  \epsilon\check{I} + \check{I}\epsilon
  ]
  (t,t)
  \,,
\end{align}
where $\check{\sigma}$ are spin matrices in Keldysh space, and
$\epsilon(t,t')=i\partial_t\delta(t-t')$ and
$\mu(t,t')=\delta(t-t')\partial_t\phi(t)$. Moreover,
we fix the potential of the right terminal to zero.
The heat currents as defined above satisfy conservation of energy at
each instant of time, $\mu(t)I(t) = \dot{Q}_L(t) + \dot{Q}_R(t)$.

To make a connection with Ref.~\onlinecite{golubev2013-htt},
it is advantageous to rewrite the heat currents in a different form,
using $e^{i\phi}\epsilon=(\epsilon+\mu)e^{i\phi}$,
\begin{align}
  \label{eq:QLdefb}
  \dot{Q}_L
  &=
  \frac{1}{8}
  \tr
  \check{\sigma_1}[\dot{g}_L\check{Z}g_R - g_R \check{Z} \dot{g}_L](t,t)
  +
  \frac{\partial_t Q_J(t)}{2}
  \,,
  \\
  \label{eq:QRdefb}
  \dot{Q}_R
  &=
  \frac{1}{8}
  \tr
  \check{\sigma_1}[\dot{g}_R\check{Z}g_L - g_L \check{Z} \dot{g}_R](t,t)
  +
  \frac{\partial_t Q_J(t)}{2}
  \,,
\end{align}
where, $\dot{g}_R=\epsilon g_{R,0}$ and
$\dot{g}_L=e^{i\phi\tau_3}\epsilon{}g_{L,0}e^{-i\phi\tau_3}$.
The quantity appearing in the second terms is
\begin{align}
  Q_J(t)
  \equiv
  i
  \frac{1}{8}
  \tr\check{\sigma}_1
  \Bigl[
  \check{g}_L \check{Z} \check{g}_R + \check{g}_R \check{Z} \check{g}_L
  \Bigr](t,t)
  \,.
\end{align}
The total time derivative $\partial_t Q_J$ does not contribute to
time-averaged heat currents.  Below, as far as heat currents are
concerned, we mostly discuss the time-averaged quantities,
$\dot{Q}_{dc,L/R}$, as these are experimentally more easily
accessible.

In the case of low-transparency tunnel junctions, we have $\check{Z}=1/R_T$,
where $R_T$ is the tunnel resistance. The approach then coincides with
that discussed in Ref.~\onlinecite{golubev2013-htt}.  Comparison to
the tunneling Hamiltonian calculation of
Ref.~\onlinecite{golubev2013-htt} identifies first terms of
Eqs.~\eqref{eq:QLdefb}, \eqref{eq:QRdefb} as the heat carried by
quasiparticles into the bulk of the terminals.  Moreover, the term
$Q_J\sim\avg{H_T(t)}$, the component of the internal
energy associated with the part of the Hamiltonian coupling the two
terminals together.  For slow phase variations, $\dot{\phi}\to0$, one
can verify that the expression is related to the Josephson energy of the
junction at equilibrium,
$\partial_tQ_J(t)\to{}I_J(t)\dot{\phi}(t)=\partial_t E_J(\phi(t))$.

One can compute the sums over $\tau_n$ for different transparency
distributions. A relevant case is the eigenvalue distribution
corresponding to a dirty interface \cite{schep97,melsen1994}
\begin{align}
  \label{eq:dirty-interface-rho}
  \rho(\tau)=\frac{h}{2\pi{}e^2R}\frac{1}{\tau^{3/2}\sqrt{1-\tau}}\,,
\end{align}
which has been experimentally found to agree with high-transparency
oxide junctions. \cite{naveh2000-udt} From this,
\begin{align}
  \label{eq:dirty-interface}
  \check{Z}
  &=
  \frac{1}{R}
  \frac{1}{
    \sqrt{2 + [\check{g}_L,\check{g}_R]_+}
  }
  \,.
\end{align}
The result for diffusive junctions in the short junction limit (dwell
time $\tau_D\ll{}\hbar/\Delta$) can be found either by solving the
Usadel equation, \cite{usadel1970-gde,artemenko1978-eci,nazarov94} or averaging
over the corresponding transparency distribution
\cite{dorokhov1984}. The former approach readily yields:
\begin{align}
  \check{I} = \frac{2}{R}\log \check{g}_L \check{g}_R
  \,.
\end{align}
As discussed in Ref.~\onlinecite{virtanen2014-fhc}, the resulting heat
current in diffusive junctions in the absence of voltage bias turns
out to be independent of the phase difference $\varphi$ over
the junction.

The above approach can also be used to compute statistics of heat
transport.  The problem can be formulated as
follows:~\cite{kindermann2004-sht,esposito2009-nff} the
internal energy $Q_R$ of one terminal is probed two times, first at
$t=0$ and a second time at $t=t_0$. The cumulant generating function
${\cal S}_R(t_0,u)$ of the energy change $\Delta Q_R=Q_R(t_0)-Q_R(0)$
can in the long-time limit ($t_0\to\infty$) be written as
\cite{kindermann2004-sht,snyman2008-kam,esposito2009-nff}
\begin{align}
  \label{eq:generating-function}
  \partial_u {\cal S}_R(t_0,u) 
  &= 
  it_0\dot{Q}_{R,dc}\rvert_{\check{g}_{R,0}\mapsto\check{g}_{R,0}(u)}
  \,,
  \\
  \check{g}_{R,0}(u)
  &\equiv
  e^{iuE\check{\sigma_1}/2}
  \check{g}_{R,0}
  e^{-iuE\check{\sigma_1}/2}
  \,.
\end{align}
The resulting statistics in the absence of driving are discussed in
Ref.~\onlinecite{virtanen2014-fhc}, and results including driving are
for the tunnel limit discussed in Ref.~\onlinecite{golubev2013-htt}.
Below, as far as fluctuations are concerned, we mainly concentrate on
the zero frequency heat noise $S_E \equiv -\partial_u^2 {\cal
  S}_R\rvert_{u=0}/t_0=\avg{(\dot{Q}_R - \avg{\dot{Q}_R})^2}$.

The above formulation is directly accessible to numerical calculations
in the matrix representation~\eqref{eq:matrixrepr}. The numerical
results below are found by keeping only a finite number of the
harmonics $k$ and truncating the energy range to energies within
several $|\Delta|, T$ from the Fermi level.  Matrix inverses,
exponentials, square roots, and logarithms can be computed efficiently
numerically.  \cite{higham2005-sas,deadman2013-bsa}

\section{Driven quasiparticle transport}
\label{sec:qp-transport}

The electronic dc heat current is carried by quasiparticles, which are
either transferred from one side to the other, or created or destroyed
in pairs by breaking or creating Cooper pairs in the
condensate. During their traversal through the junction area, the
quasiparticles can either gain or lose energy via interaction with the
electromagnetic field.

Within the formalism described here, the dc heat current entering terminal
$\alpha=L/R$ can in general be written in terms of the quasiparticle
distributions in the two terminals:
\begin{align}
  \label{eq:heat-flow-Y}
  \dot{Q}_{dc,\alpha}
  &=
  \frac{1}{R}
  \int_{-\infty}^\infty\dd{E}
  E
  \sum_{\beta=L,R}
  \sum_{k=-\infty}^\infty
  \\\notag
  &\quad
  \times
  Y_{\alpha,\beta,k}(E)\Bigl[\tanh\bigl(\frac{E + k\omega_0}{2T_\beta}\bigr)-\tanh\bigl(\frac{E}{2T_\alpha}\bigr)\Bigr]
  \,.
\end{align}
Expressions for $Y_{\alpha\beta{}k}(E)$ can be found with some
algebraic manipulations discussed in Appendix~\ref{sec:manipulations}.
The functions $Y_{\alpha,\beta,k}(E)$ can be understood to be
proportional to the total rates of processes in which a quasiparticle
starting at energy $E+k\omega_0$ in terminal $\beta$ ends up in
terminal $\alpha$ at energy $E$ while losing or gaining energy
$k\omega_0$ by interacting with the electromagnetic field in the
junction.

One should note that the rates $Y$ do not correspond to elementary
energy transfer processes in the counting statistics sense.  Rather,
they are simply proportional to time-averaged total rates.  Analogous
expressions as Eq.~\eqref{eq:heat-flow-Y} for the heat current can in
normal-state systems be written in terms of the Floquet scattering
matrices. \cite{moskalets2014-fsm,battista2014-eap}

The total power dissipated in the junction is
\begin{align}
  P = \dot{Q}_{dc,L} + \dot{Q}_{dc,R}
  \,.
\end{align}
In general, it is directly connected to the charge current in the
junction, by conservation of work done by the bias, $P=\overline{IV}$.
For the discussion below, we also define the temperature-driven heat
current as
\begin{align}
  {\cal J}=(\dot{Q}_{dc,L} - \dot{Q}_{dc,R})/2
  \,.
\end{align}
The result is a well-defined temperature-driven current provided the
dissipation is low, $P\ll{}|{\cal J}|$, which requires a large enough
temperature difference between the two sides of the junction.

\subsection{Dc voltage bias}
\label{sec:voltage-bias}

\begin{figure}
  \includegraphics{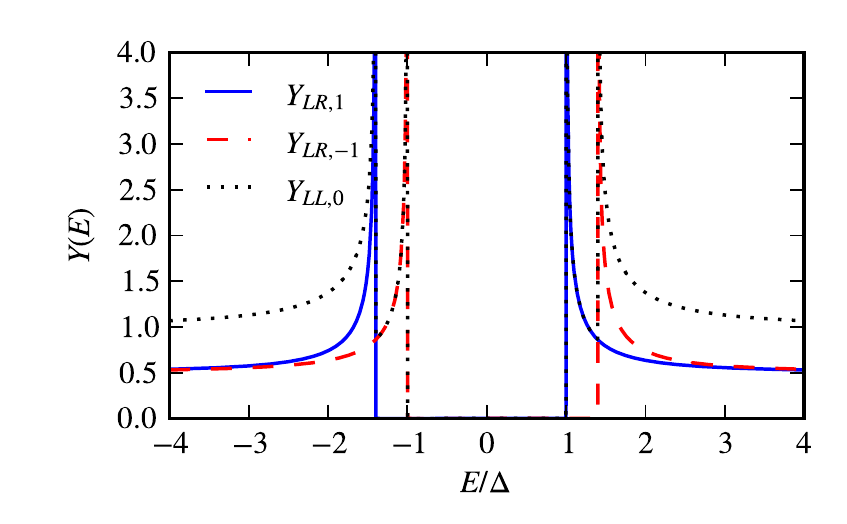}
  \caption{
    \label{fig:tunnel-Y-V}
    \emph{(Color online)}
    Quasiparticle transport factors $Y_{LL,0}(E)$ (dotted) and $Y_{LR,\pm1}(E)$ (solid and dashed)
    for tunnel junction at constant bias voltage, $\phi(t)=Vt$, $V=0.4\Delta$.
    Analytic results in Eq.~\eqref{eq:tunnelling-Y} are shown in dotted lines.
  }
\end{figure}

\begin{figure}
  \includegraphics{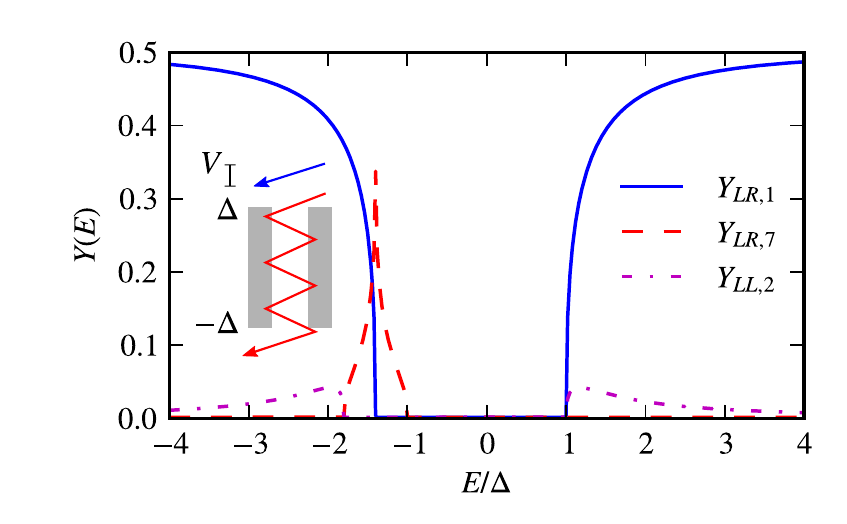}
  \caption{
    \label{fig:qpc-Y-V}
    \emph{(Color online)} Energy flow functions $Y_{\alpha\beta,k}(E)$
    for quantum point contact with $\tau=1$ at constant bias voltage,
    $\phi(t)=Vt$, $V=0.4\Delta$. Only processes with energy absorption
    are shown. Inset: schematic energy diagram
    showing the quasiparticle trajectories corresponding to $Y_{LR,1}$
    (direct quasiparticle transport) and $Y_{LR,7}$ (MAR). Note that
    Andreev reflection probability is finite also at $|E|>\Delta$.  }
\end{figure}

In dc voltage biased junctions with a finite transparency, the
well-known multiple Andreev reflection processes (MAR)
\cite{tinkham1996-its,octavio1983-ses} play a role in the
quasiparticle transport.  They lead to diffusion of quasiparticles
upward in energy inside the junction, and in general give rise to
several side bands $Y_{LR,k\ne0}$, corresponding to quasiparticles
Andreev reflected several times inside the junction.

Similarly as for the charge transport
\cite{cuevas2004-dti,octavio1983-ses,cuevas1996-hat}, the calculation
of the heat transport and its fluctuations in the dc voltage biased case
reduces to solving a relatively simple recurrence equation, describing
the propagation of quasiparticles upwards in energy. We defer the
derivation of the equations to Appendix~\ref{sec:dc-recurrence}. They
can be solved analytically in certain limiting cases, and are in
general straightforward to solve numerically.

The tunnel junction limit is one of the analytically tractable cases,
and for it one finds \cite{golubev2013-htt,guttman1997-pdt}
\begin{align}
  \label{eq:tunnelling-Y}
  Y_{L,R,k} &= N_L(E)[\delta_{k,1}N_R(E+V) + \delta_{k,-1} N_R(E-V)]
  \,,
  \\
  Y_{L,L,k} &= [Y_{L,R,1} + Y_{L,R,-1}]\delta_{k,0}
  \,,
\end{align}
where $N_{L/R}$ are the densities of states of the
superconductors. The result is shown in Fig.~\ref{fig:tunnel-Y-V} for
reference. The resulting expression for the electronic heat current is
similar to that in normal-state junctions, with the superconductivity
only modifying the quasiparticle densities of states of the two
terminals.

In junctions with finite transparency, Andreev reflections as expected
start to facilitate quasiparticle transport across the energy gap
region. The corresponding factors $Y_k$ are plotted in
Fig.~\ref{fig:qpc-Y-V} for a fully transparent junction
($\tau=1$). One can note that in addition to the direct quasiparticle
transmissions, one obtains MAR processes for $|k|>2\Delta/V$.

\begin{figure}
  \includegraphics{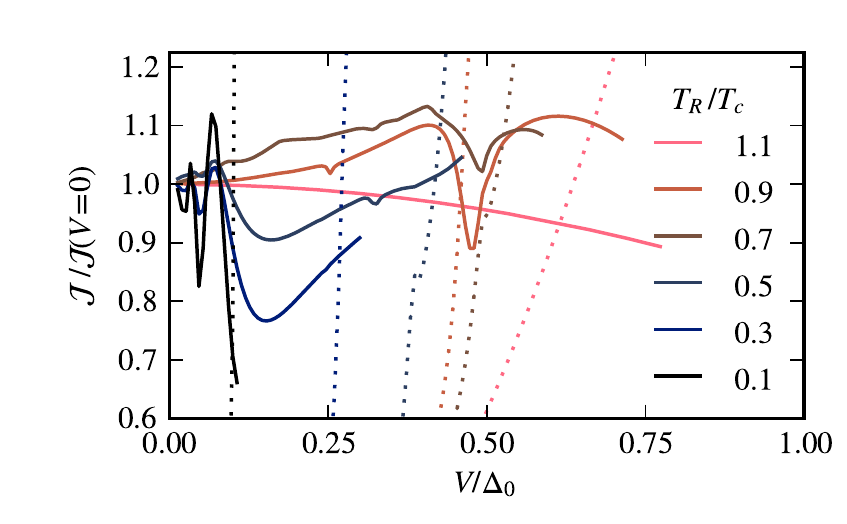}
  \caption{ 
    \label{fig:qpc-heat-mar}
    Heat current ${\cal J}$ (solid lines) through a quantum point
    contact with $\tau=0.5$, as a function of bias voltage, normalized by its
    zero-bias value.  Temperature difference is present, with $T_L=0$
    and $T_R$ varying.  Dotted lines indicate the normalized power
    $\overline{I V}/{\cal J}(V=0)$ dissipated in the junction.
    Results are plotted up to voltages where the dissipated power $P$
    becomes larger than the heat current.
  }
\end{figure}

Results for the heat current under the simultaneous application of a
temperature difference and bias voltage are shown in
Fig.~\ref{fig:qpc-heat-mar}, corresponding to a quantum point contact
with partial transparency.  One can note that features of multiple
Andreev reflections do not appear to play a large role in the
temperature-driven heat current as long as $P\ll{}{\cal J}$.  This can
be seen explicitly by considering the contributions in
Eq.~\eqref{eq:heat-flow-Y} separately for each sideband $k$ (not
shown), which indicates direct quasiparticle transfer ($Y_{k=\pm1}$)
is responsible for the low-bias features. At higher bias when
$P\gtrsim{}{\cal J}$, especially in the high-temperature curves for
which $\Delta_R(T_R)<\Delta_L(T_L)$, MAR processes do not contribute
in the same way to $\dot{Q}_R$ and $\dot{Q}_L$. This results to sharp
changes in ${\cal J}=(\dot{Q}_R-\dot{Q}_L)/2$ at voltages close to
thresholds of the different possible MAR processes.

\begin{figure}
  \includegraphics{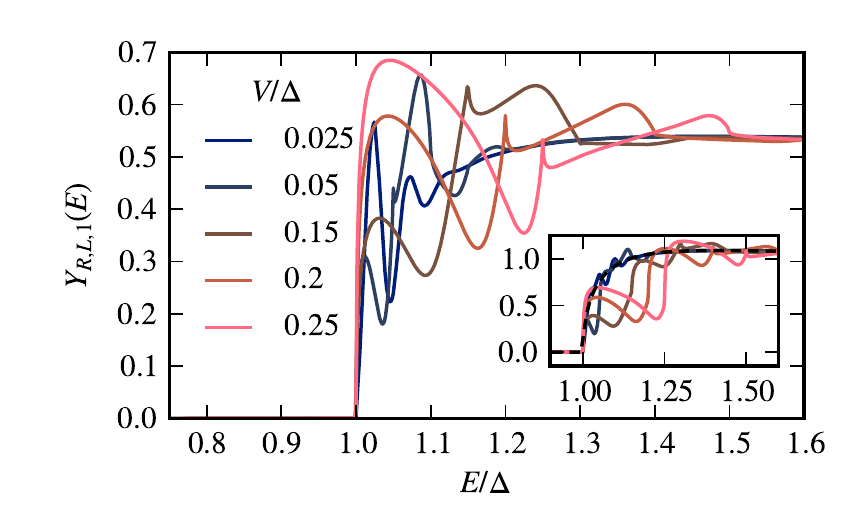}
  \caption{
    \label{fig:qpc-Y-V-dep}
    Bias dependence of $Y_{LR,1}$ for $\tau=0.5$.
    The functions are symmetric around $E=-V/2$.
    The reduction of transport at energies $\Delta<E<\Delta+V$
    for intermediate $V/\Delta\sim0.15$
    is responsible for the reduction of heat current in
    Fig.~\ref{fig:qpc-heat-mar}.
    Inset: Bias dependence of $\sum_k Y_{LR,k}$. For $V\to0$,
    the result converges towards the adiabatic limit (dashed)
    given by averaging the dc result over phase differences $\varphi_0$
    [Eq.~\eqref{eq:tau05-zerobias}].
  }
\end{figure}

The nonmonotonic behavior at low voltages in
Fig.~\ref{fig:qpc-heat-mar} arises from modification of the resonance
in $Y_{LR,\pm1}$ at $|E|>\Delta$.  The evolution of the resonance in
$Y$ with increasing bias voltage is shown in
Fig.~\ref{fig:qpc-Y-V-dep}. Initially, transport at energies
immediately above the gap is reduced. A somewhat similar effect also
occurs in the tunnel junctions, as described by
Eq.~\eqref{eq:tunnelling-Y}, where the peak in $Y_{LR,1}$ near
$E=\Delta$ is reduced from $N(E)^2$ to $N(E)N(E+V)$.  The behavior of
partially transparent junctions differs from tunnel junctions
in that the amplitude of the resonance increases again at larger
voltages, resulting to a non-monotonic behavior in the heat current.
The effect is the largest at low temperatures, as visible in
Fig.~\ref{fig:qpc-heat-mar}, as transport close to gap edges has the
most importance in that case.

\begin{figure}
  \includegraphics{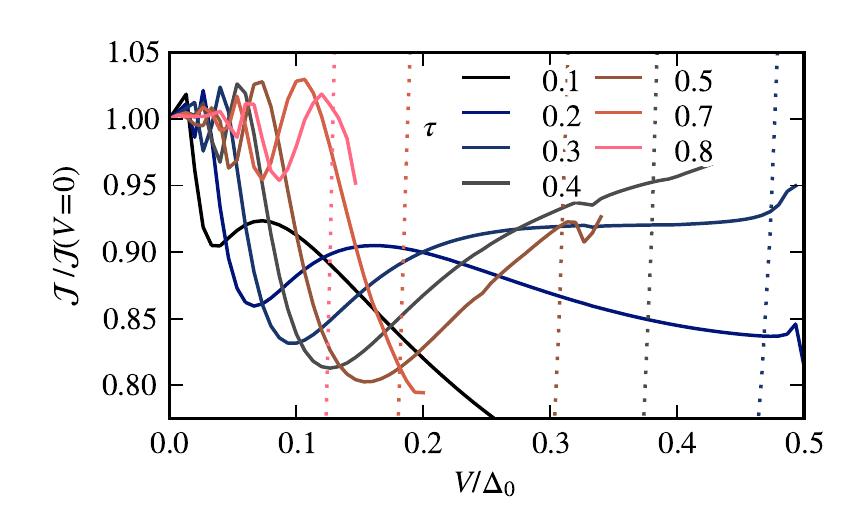}
  \caption{ 
    \label{fig:qpc-heat-mar-tau}
    Heat current ${\cal J}$ in a QPC as a function of bias, for fixed
    $T_L=0$ and $T_R=0.35T_c$ and different $\tau$.  The
    adiabatic limit result ${\cal J} = \overline{{\cal
        J}(\varphi(t))}$ is indicated with dots on the left axis, and
    the dissipated power $\overline{I V}$ is shown with dotted lines.
  }
\end{figure}

Figure~\ref{fig:qpc-heat-mar-tau} displays the dependence of the heat
current on the bias voltage for different values of the transparency.
The above nonmonotonic behavior becomes more prominent as the junction
transparency increases and the suppression of heat current is lifted.
As expected, MAR processes also become more significant as the
transparency increases: the higher the transparency, the lower the
bias voltage $V$ at which the dissipated power $P$ overwhelms the heat
current ${\cal J}$.

\begin{figure}
  \includegraphics{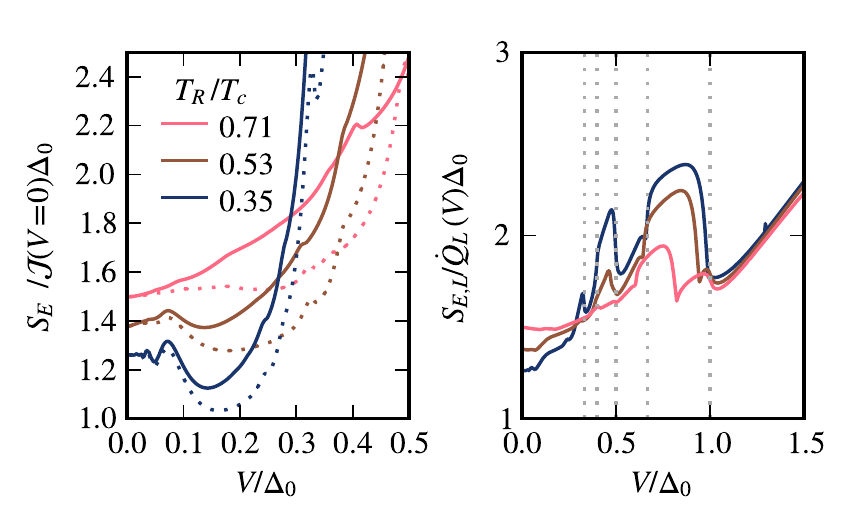}
  \caption{
    \label{fig:qpc-heat-mar-and-noise}
    Heat current noise $S_E$ as a function of bias voltage $V$ for a
    point contact with $\tau=0.5$. Noise in the heat current into left
    ($S_{E,L}$, solid) and right ($S_{E,R}$, dotted) terminals is
    shown, for different temperatures $T_R$ keeping $T_L=0$ fixed.
    Right panel shows the noise normalized to the heat current.
    Zero temperature MAR thresholds $V=2\Delta_0/n$ are indicated
    with dotted lines.
  }
\end{figure}

How multiple Andreev processes contribute to the heat transport is
also visible in the heat transport noise, shown in
Fig.~\ref{fig:qpc-heat-mar-and-noise}. At low bias, the voltage
dependence of the noise follows that of the heat current itself,
a feature similar to what occurs in a stationary situation without
bias. \cite{virtanen2014-fhc} At high bias, the noise increases
rapidly as the MAR processes activate. The MAR features are best visible in the
ratio $S_E/\dot{Q}$, which has sharp features at the activation
thresholds $V=2\Delta/n$. For a Poissonian process of absorbing energy
packets $E_0$, the ratio $S_E/\dot{Q}=E_0$ indicates directly the
energy transferred by a single elementary process. Although the
situation in the present case is more complicated, the fact that order
$n=\mathrm{ceil}(2\Delta/V)$ MAR process is associated with absorption of
energy $E_0\sim{}nV\sim{}\mathrm{ceil}(2\Delta/V)V$ results to
oscillatory behavior similar as seen in
Fig.~\ref{fig:qpc-heat-mar-and-noise} (right panel).

\begin{figure}
  \includegraphics{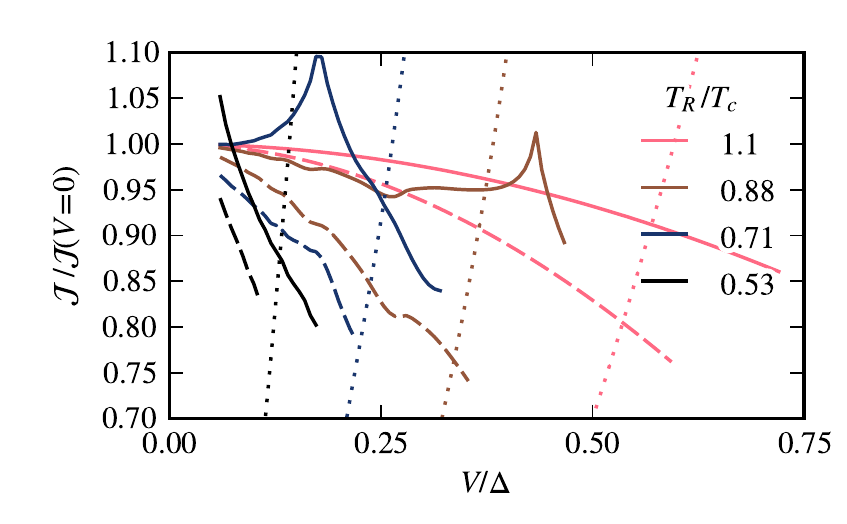}
  \caption{ 
    \label{fig:heat-mar-dirty-and-diffusive}
    As Fig.~\ref{fig:qpc-heat-mar}, but for the dirty interface 
    (solid ${\cal J}$, dotted $P$)
    and diffusive distributions (dashed ${\cal J}$) of transmission channels.
  }
\end{figure}

Finally, Fig.~\ref{fig:heat-mar-dirty-and-diffusive} shows the bias
dependence of the temperature-driven heat current for the dirty
interface and diffusive channel distributions. The behavior of the
heat current in these cases is dominated by the large number of
low-transparency channels, and is mainly similar to low-transparency
junctions in that the heat current decreases with increasing bias
without significant oscillations.  The peaks appearing in the dirty
interface case occur at $V=\Delta_L(T_L)-\Delta_R(T_R)$, and are due
to matching the peaks in the above-gap DOS in the superconductors.
A similar feature is present also in tunnel junctions. \cite{golubev2013-htt}
In contrast, the results for the diffusive distribution contain less
low-transparency channels, and the gap-difference features are not
clearly resolved.

\subsection{Ac phase bias}
\label{sec:ac-bias}

Consider now the case of a junction under monochromatic ac phase
bias, $\phi(t) = \varphi_0/2 + 2s\cos(\omega t)$, where
$s=V_{ac}/(2\hbar\omega)$.  Such an excitation in general has two
effects --- first, the microwave photons $\hbar\omega$ can be
absorbed, which can result to dissipation in the form of
photon-assisted tunneling and breaking of Cooper pairs. Second, the
time dependent phase modulates the coherent
transmission resonance responsible for carrying a significant portion
of the heat current.

\begin{figure}
  \includegraphics{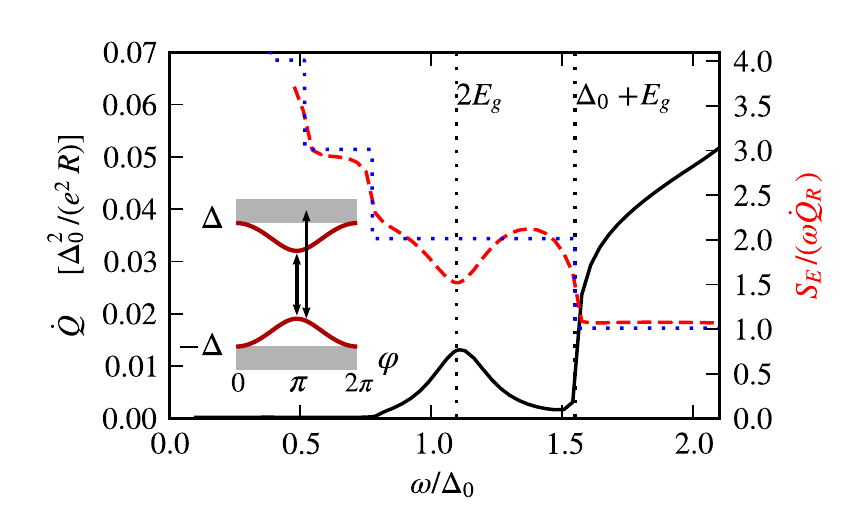}
  \caption{
    \label{fig:ABS-absorption}
    Dissipated power (solid, left axis) and heat transport noise (dashed, right axis)
    in a QPC with $\tau=0.7$ under ac drive
    $\phi(t)=\pi/2 + 0.2\cos(\omega t)$ with $T_L=T_R=0$.
    The frequency thresholds
    associated with pair breaking facilitated by ABS
    are clearly visible.
    Noise $S_E=n\omega\dot{Q}_R$ expected from Poissonian
    $n=\lceil(\Delta+E_A)/\omega\rceil$ photon statistics is also shown (dotted).
    Inset: illustration of the processes corresponding to
    the thresholds shown.
  }
\end{figure}

\begin{figure}
  \includegraphics{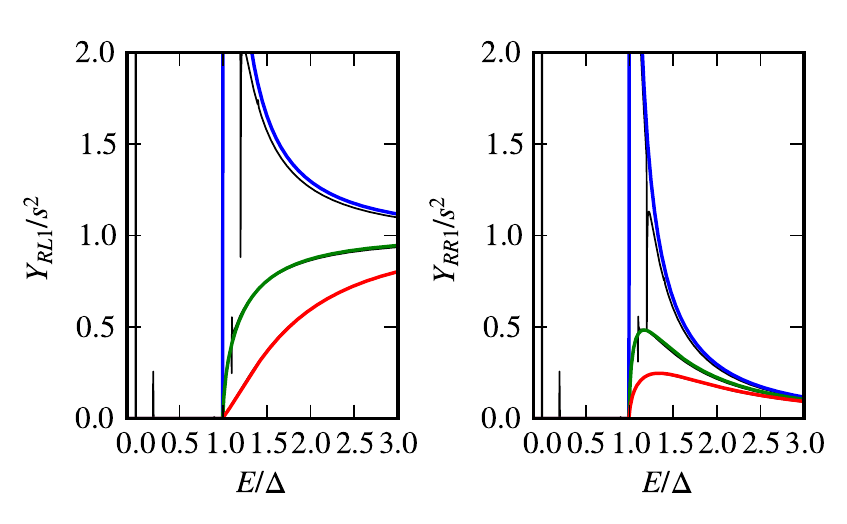}
  \caption{
    \label{fig:qpc-1-Y-an}
    $Y_{R,L/R,+1}$ related to one-photon absorption processes, for a
    fully transparent QPC ($\tau=1$) under harmonic drive
    $\phi(t)=\varphi_0/2 + 2s\cos(\omega t)$.  Here,
    $\omega=\Delta/5$, and the result is shown for
    $\varphi_0=0,\frac{\pi}{2},\pi$ (curves from top to bottom).  
    The functions are symmetric around the point $E=-\omega/2$.
    The results are from Eqs.~\eqref{eq:ac-phase-y-1}, \eqref{eq:ac-phase-y-2}, which
    are valid for $s\to0$.
    Thin black lines indicate results from Eq.~\eqref{eq:y-functions} for $s=0.1$,
    which have additional multiphoton resonances due to bound states.
  }
\end{figure}

Let us first discuss the dissipation in the junction, for which
Cooper-pair breaking transitions from negative to positive energies
are important. In particular, the Andreev bound states are crucial for
this, as they reduce the frequency threshold for Cooper-pair
breaking. \cite{shumeiko1993-res,gorelik1995-mis,kos2013-fda} An ABS
at energy $E_A<\Delta$ enables pair breaking by 1-photon processes to
occur via processes involving only the bound states at $\omega=2E_A$
and via processes involving also the continuum at $\omega>\Delta+E_A$,
as also seen e.g. in Refs.~\onlinecite{bergeret2010-tom,kos2013-fda}.
These frequency thresholds occur prominently in the dissipated power
shown in Fig.~\ref{fig:ABS-absorption}.  \footnote{ The matrix element
  for the $2E_A$ process is $\propto\sqrt{1-\tau}$,
  \cite{zazunov2003-alq} so that the feature is absent in
  Fig.~\ref{fig:qpc-1-Y-an}.} Analytical results for the dissipated
power in the representative case $\Delta_L=\Delta_R=\Delta$,
$T_L=T_R=T$ in the limit $s\to0$ can be deduced from results in
Ref.~\onlinecite{kos2013-fda}.

Figure~\ref{fig:ABS-absorption} also displays the heat transport noise,
computed numerically via Eq.~\eqref{eq:generating-function}.  At high
frequencies, the result converges towards $S_E\to\omega\dot{Q}_R$, and
one can check that also the third cumulant $\partial_u^3{\cal
  S}_R(u)\rvert_{u=0}\approx{}it_0\omega^2\dot{Q}_R$ in this
limit. That the cumulants coincide suggests that photon absorption
follows Poisson-like statistics. The situation at lower frequencies
$\omega<\Delta+E_A$ is somewhat more complicated, as multiple photons
are required for each absorption event.  For Poissonian multiphoton
processes, we expect $S_E=n\omega\dot{Q}_R$ for
$(E_A+\Delta)/n<\omega<(E_A+\Delta)/(n-1)$. This is in line with the
results in Fig.~\ref{fig:ABS-absorption}, apart from features that
come from the second possible transition process involving only the
bound states.

At frequencies below the transition thresholds, dissipation by photon
absorption via processes involving the ABS become less
important. Rather, the main physics comes from quasiparticles in the
continuum and from the modification of their phase dependent transport
resonances.  Analytical results can be obtained by taking the drive
amplitude $s$ as a small parameter. Working to order $s^2$ and
considering only processes in the continuum ($E>\Delta$), we find for
a fully transparent QPC ($\tau=1$),
\begin{align}
  \label{eq:ac-phase-y-1}
  Y_{RL,\pm1}
  &=
  s^2
  \frac{
    2B_0 B_{\pm\omega}
    [
    (E\pm\omega)E
    +
    \Delta^2\cos\varphi_0
    +
    B_0 B_{\pm\omega}
    ]
  }{
    A_0 A_{\pm\omega}
  }
  \,,
  \\
  \label{eq:ac-phase-y-2}
  Y_{RR,\pm1}
  &=
  s^2
  \frac{
    2B_0 B_{\pm\omega}
    [
    (E\pm\omega)E
    +
    \Delta^2
    -
    B_0 B_{\pm\omega}
    ]
  }{
    A_0 A_{\pm\omega}
  }
  \\
  \label{eq:ac-phase-y-3}
  Y_{RL,0}
  &=
  \theta_0
  \frac{2(E^2 - \Delta^2)}{A_0}
  \left(
    1
    -
    s^2
    \sum_\pm
    \frac{
      C_{\pm\omega}
    }{
      A_0 A_{\pm\omega}
    }
    \right)
    \,,
  \\\notag
  A_\omega &= 2(E+\omega)^2 - (1 + \cos\varphi_0)\Delta^2
  \,,
  \\\notag
  B_\omega &= \theta_{\omega}\sqrt{(E+\omega)^2 - \Delta^2}
  \,,
  \quad
  \theta_\omega=\theta((E+\omega)^2-\Delta^2)
  \,,
  \\\notag
  C_\omega &=
  [
  2B_0B_{\omega}-(1-\cos\varphi_0)\Delta^2
  ]
  \\\notag
  &\quad\times
  [2E(E+\omega) + (1+\cos\varphi_0)\Delta^2]
  \,.
\end{align}
The divergence in $Y_{RL,0}$ at $|E|=\omega+\Delta|\cos(\varphi_0/2)|$
is to be understood in the principal value sense --- the above result
does not describe the dissipative processes involving the Andreev
bound states.  Details of the calculation and results for $\tau\ne1$
are in Appendix~\ref{sec:ac-bias-expansion}. In the tunneling limit,
$\tau\to0$, on the other hand,
\begin{align}
  Y_{RL,\pm1}
  &=
  \theta_0\theta_{\pm\omega}
  \frac{
    (E\pm\omega)E + \Delta^2\cos\varphi_0
  }{
    \sqrt{E^2-\Delta^2}
    \sqrt{(E\pm\omega)^2-\Delta^2}
  }
  \,,
  \\
  Y_{RR,\pm1}
  &\approx0
  \,,
  \\
  \label{eq:tunnelling-YRL0}
  Y_{RL,0}
  &=
  \theta_0
  \frac{
    E^2 - \Delta^2\cos\varphi_0
  }{
    E^2 - \Delta^2
  }
  (1 - 2s^2)
  \,.
\end{align}
The results for $\tau=1$ are shown in Fig.~\ref{fig:qpc-1-Y-an}.  The
energy dependence of the driven response reflects the behavior in the
stationary state: there is a resonant contribution at energies
immediately above the superconducting gap, which is for transparent
junctions reduced when the Andreev bound states at
$E_A=\sqrt{1-\tau\sin^2(\varphi/2)}$ separate from the gap edges.

\begin{figure}
  \includegraphics{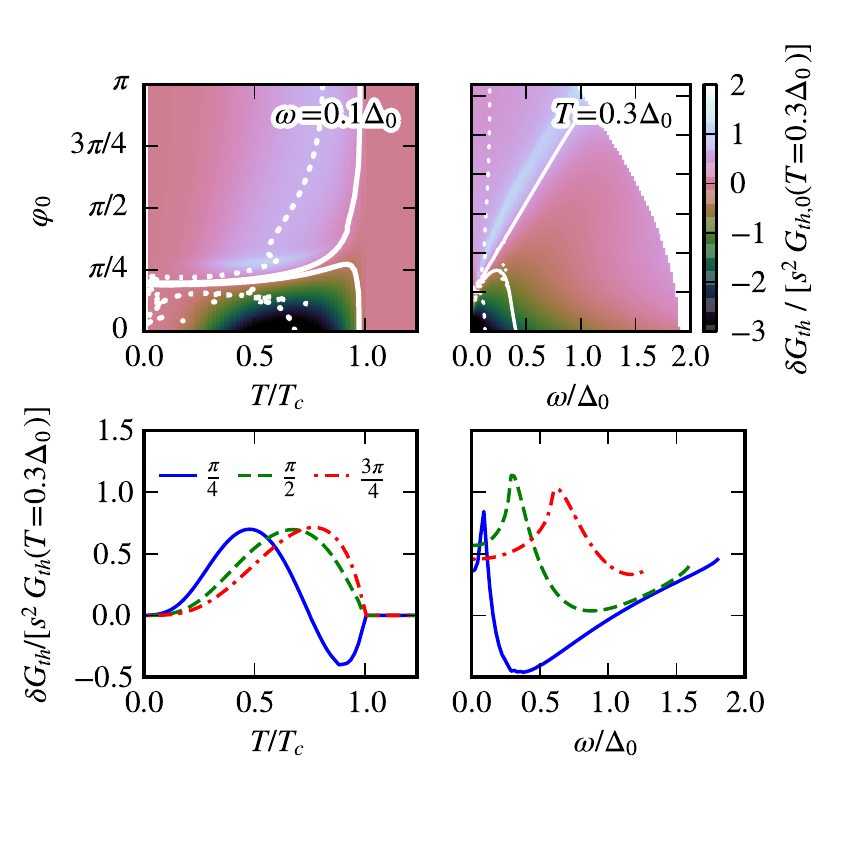}
  \caption{
    \label{fig:Gth-change-tau1}
    Change of heat conductivity in a $\tau=1$ QPC caused by small
    harmonic drive, $s>0$. Shown as a function of phase, temperature
    (left panels) and frequency of the drive (right panels). The contour
    lines denote regions where change in dissipation is $>10\%$ (dotted)
    and $>100\%$ (solid) of the change in the temperature-driven heat
    current for $T_R-T_L=0.1\Delta$. The region $\omega>\Delta+E_A$ where 1-photon Cooper pair
    breaking activates is omitted in the plots, as it is not described by
    Eqs.~\eqref{eq:ac-phase-y-1}--\eqref{eq:ac-phase-y-3}.
  }
\end{figure}

\begin{figure}
  \includegraphics{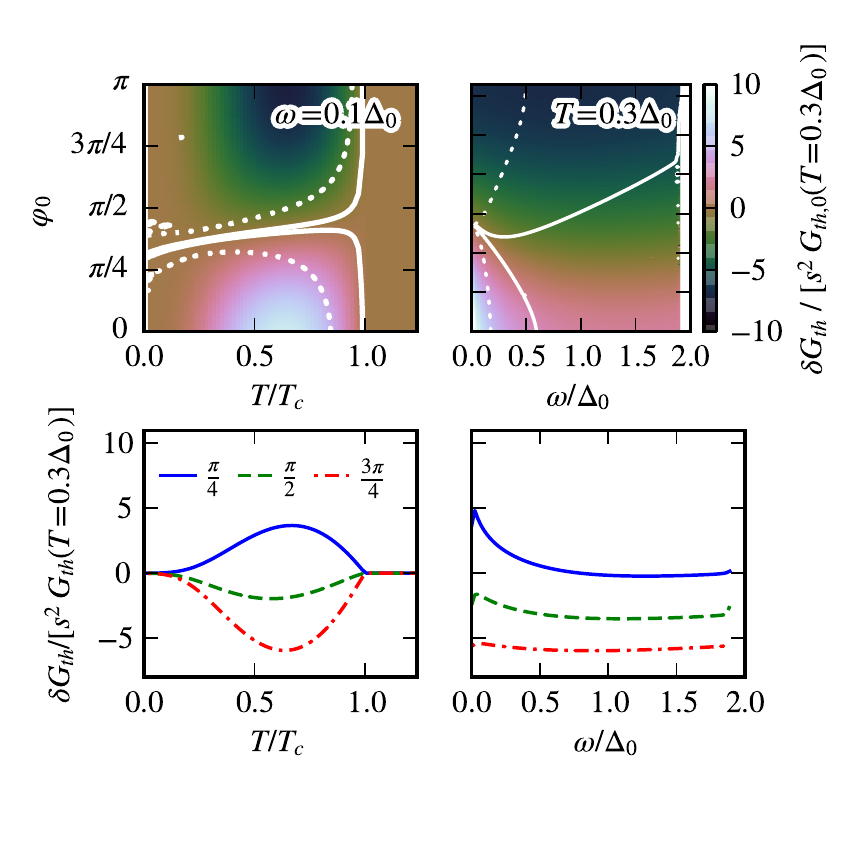}
  \caption{
    \label{fig:Gth-change-tau0}
    As Fig.~\ref{fig:Gth-change-tau1}, but for a low-transparency
    junction $\tau=0.01$.
  }
\end{figure}

The resulting change $\delta G_{th} = G_{th}(s) - G_{th}(s=0)$ in the
heat conductivity $G_{th}={\cal J}/(T_L-T_R)$ in the limit $s\to0$ is
shown in Fig.~\ref{fig:Gth-change-tau1} for a transparent junction and
in Fig.~\ref{fig:Gth-change-tau0} for a low-transparency junction. For
the transparent junction case, the effect of the ac excitation is the
largest at phase differences $\varphi_0\approx0$, as expected on the
basis of Fig.~\ref{fig:qpc-1-Y-an}. Moreover, the excitation tends to
suppress the heat flow more than it enhances it.  The dependence on
$\varphi_0$ in the tunneling limit is opposite to the transparent
limit, reflecting the similarly reversed phase dependence in the
stationary case \cite{zhao2003-phase}.  In the tunneling limit, the
resonance has the largest effect at $\varphi_0\approx\pi$, and the ac
drive reduces its contribution to the heat current.

The amount of power dissipated relative to the change in heat current
for a $0.1\Delta_0$ temperature difference is also indicated in
Figs.~\ref{fig:Gth-change-tau1}-\ref{fig:Gth-change-tau0} with contour
lines. In the regime shown, power dissipation is mainly due to processes at
$E>|\Delta|$, which are well described by
Eqs.~\eqref{eq:ac-phase-y-1}--\eqref{eq:ac-phase-y-3}. The features
seen in the left panels mainly come from the fact that the change 
$\delta G_{th}$ reverses its sign around $\varphi_0\approx\pi/4$ both for
tunnel and transparent junctions --- as seen in
Figs.~\ref{fig:Gth-change-tau1}-\ref{fig:Gth-change-tau0}, the sign
reversal is only weakly temperature and frequency dependent.  The main
feature of the dissipation in the right panel is that if the phase
excitation amplitude is kept fixed, the dissipated power grows with
increasing frequency, whereas the effect of the drive on the
temperature-driven heat current saturates towards higher
frequencies. Moreover, at sufficiently high frequency, the Cooper
pair breaking processes activate, and start to dominate the heat flows.

\begin{figure}
  \includegraphics{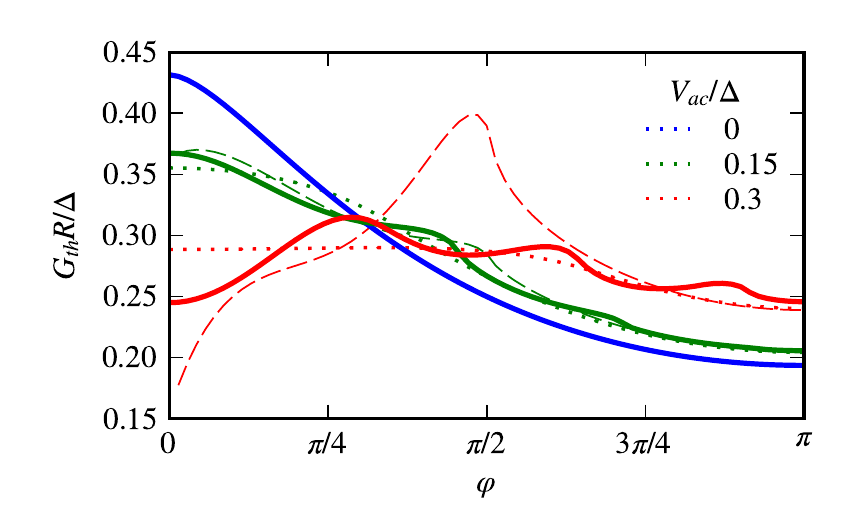}
  \caption{
    \label{fig:Gthphaseosc-qpc1w03}
    Phase oscillations of heat conductivity in a QPC with $\tau=1$,
    at $T/\Delta=0.3$ and $\omega/\Delta=0.3$, for different
    drive amplitudes.
    Results from Eqs.~\eqref{eq:ac-phase-y-1}--\eqref{eq:ac-phase-y-1} (dashed)
    and adiabatic limit (dotted) are also shown.
  }
\end{figure}

Phase oscillation of heat conductivity as a function of increasing
drive amplitude is shown in Fig.~\ref{fig:Gthphaseosc-qpc1w03}.
Initially, the response is well described by the leading-order theory,
but starts to deviate from the $s\to0$ limit when the drive amplitude
exceeds $s\gtrsim0.1$. The result moreover stays close to the
adiabatic limit $\overline{G_{th}(\varphi(t))}$, obtained by averaging the
stationary heat-current--phase relation over the drive cycle.

\begin{figure}
  \includegraphics{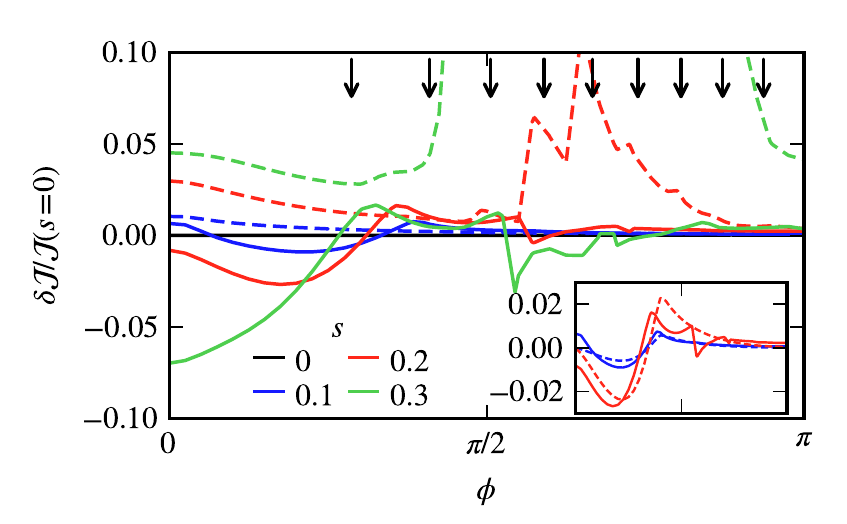}
  \caption{
    \label{fig:Jth-diffusive-osc}
    Emergence of phase-dependent oscillations in the heat current
    through a short diffusive junction, when it is driven by ac bias
    $\phi(t)=\varphi_0/2 + 2 s \cos(\omega t)$, with $\omega
    /\Delta=0.2$, for $T_L=0$, $T_R=0.2\Delta$.  Relative change in
    the heat current $\delta{\cal J}/{\cal J}=[{\cal J}(s)/{\cal
      J}(s=0)]-1$ (solid) and the relative dissipated power $P/{\cal
      J}(s=0)$ (dashed) are shown.  The $n$-photon process activation
    thresholds~\eqref{eq:diffusive-activation-threshold} are indicated
    with arrows.
    Inset: Comparison to results for ${\cal J}$ from
    Eqs.~\eqref{eq:general-ac-analytic-1}--\eqref{eq:general-ac-analytic-n}
    (dotted). 
  }
\end{figure}

The heat current in diffusive short junctions is independent of the
phase difference $\varphi_0$. \cite{virtanen2014-fhc} Nevertheless,
the response of the heat current to a small ac excitation is phase
dependent.  This is illustrated in Fig.~\ref{fig:Jth-diffusive-osc},
which shows the temperature-driven heat current ${\cal
  J}=(\dot{Q}_R-\dot{Q}_L)/2$ as a function of $\varphi_0$. The inset
of Fig.~\ref{fig:Jth-diffusive-osc} displays the comparison to results
from
Eqs.~\eqref{eq:general-ac-analytic-1}--\eqref{eq:general-ac-analytic-n},
which neglect transitions involving the sub-gap Andreev bound states
in the junction, but still capture the main features. The changes in
${\cal J}$ are however not very large, and are of a similar order of
magnitude or smaller compared to the power dissipated in the
junction.

The power dissipated in a diffusive junction grows rapidly at phase
differences away from zero. Such behavior is expected due to
(multi)photon absorption processes involving the sub-gap bound states,
as these processes are limited by the presence of a phase-dependent
energy gap $2E_g=2|\Delta||\cos(\varphi/2)|$ in the density of states
inside the junction.  \cite{kos2013-fda} As $\varphi$ increases, the
energy gap decreases, and $n$-photon processes activate after
corresponding thresholds are crossed:
\begin{align}
  \label{eq:diffusive-activation-threshold}
  \varphi > \varphi_n = 2\arccos\frac{n\omega}{2|\Delta|}
  \,.
\end{align}
The onsets of dissipation in the numerical results in
Fig.~\ref{fig:Jth-diffusive-osc} align with these thresholds, even
though the phase oscillation amplitude $4s$ is not very small compared
to $\pi$.  Note that in contrast to the above-gap heat transport, the
sub-gap absorption physics can be sensitive to electron-phonon and
other inelastic relaxation processes,
\cite{zazunov2005-dap,olivares2014-dqt} as excited sub-gap quasiparticles
cannot easily escape to the leads. Such mechanisms are neglected in
the above results, and the sub-gap populations are determined via
the drive-induced multiphoton coupling to the above-gap states.

\section{Discussion}
\label{sec:discussion}

The effect of electromagnetic driving on heat transport in
superconducting junctions is twofold: first, quasiparticles can absorb
energy from the field and the field can break Cooper pairs, both of
which leads to dissipation.  Second, a nonzero electromagnetic field
inside the junction implies a time dependence in the the phase
difference in superconducting order parameters. This results to
modulation and averaging of phase coherent transport effects,
which does not directly dissipate power in the junction.

In general, we find that at low drive frequencies or voltages, there
is a finite parameter range in which the response of the system is
non-adiabatic while the power dissipated by the electromagnetic fields
is still small compared to the temperature-driven heat currents.  That
is, time-dependent effects associated with modification of the
transport resonances of the continuum quasiparticles can be seen
before the dissipation heats up the system significantly.

The steady-state effects on the temperature-driven heat current ${\cal
  J}$ can in principle be measured with the help of established
experimental thermometry techniques \cite{giazotto2006-omi} for
studying heat in mesoscopic structures, similar to those applicable to
a stationary situation. In addition, the energy absorption $P$ is
experimentally accessible via certain spectroscopic approaches, used
for quantum point contacts in Ref.~\onlinecite{bretheau2013-ssa} and
it can also be deduced from susceptibility measurements, used in
Ref.~\onlinecite{chiodi2011-pdo} for diffusive junctions.

In summary, we consider heat current driven by a temperature
difference across superconducting junctions of different
transparencies.  We obtain analytical and numerical results for the
transport rates describing transport combined with energy absorption
of $k$ photons from the electromagnetic field. The results are
applicable to quantum point contacts and to generic quantum coherent
multichannel junctions.

\acknowledgments

P.V. acknowledges the Academy of Finland for financial support.  The
work of F.G. has been partially funded by the European Research
Council under the European Union's Seventh Framework Programme
(FP7/2007-2013)/ERC grant agreement No. 615187-COMANCHE, and by the
Marie Curie Initial Training Action (ITN) Q-NET 264034.

\appendix

\section{Symmetries}

The charge and heat currents in this problem have specific symmetries
with respect to transformations of the input drive $\phi$.

We can first observe that since
$-\tau_3\tau_2\check{g}_0\tau_2\tau_3=\check{g}_0$ and
$\tau_2\tau_3\tau_2=-\tau_3$, it follows that
\begin{align}
  \dot{Q}_{dc,L/R}[\phi]=\dot{Q}_{dc,L/R}[-\phi]
  \,,
  \quad
  I_{dc}[\phi]=-I_{dc}[-\phi]
  \,,
\end{align}
which is associated with the electron-hole symmetry within the model.
Moreover,
\begin{align}
  Q_{dc}[\phi]
  =
  Q_{dc}[-\phi]
  \rvert_{L\leftrightarrow{}R}
  ,
\end{align}
from gauge choice. In symmetric junctions for $T_L=T_R$, this implies
$Q_{dc,L/R}=\frac{1}{2}\overline{I V}$.

\section{Algebraic manipulations}

\label{sec:manipulations}

It is useful to start by splitting to partial fractions,
(cf. Ref.~\onlinecite{nazarov99})
\begin{align}
  \check{I}
  &=
  f(\check{g}_L \check{g}_R)
  -
  f(\check{g}_R \check{g}_L)
  =
  \tilde{f}(\check{y})
  \,,
  \\
  f(y) 
  &= 
  -4
  \sum_n
  \frac{
    q_n
  }{
    y + q_n
  }
  \,,
  \quad
  q_n = -1 + \frac{2}{\tau_n} + \frac{2\sqrt{1-\tau_n}}{\tau_n}
  \,,
\end{align}
where $\check{y} = \check{g}_L\check{g}_R$ and
$\check{y}^{-1}=(\check{g}_L
\check{g}_R)^{-1}=\check{g}_R\check{g}_L$.
The quantities $q_n$ are eigenvalues of the hermitian square
of the transfer matrix of the junction. \cite{beenakker97,nazarov99}
For continuous transmission distributions, we can define the
corresponding density
\begin{align}
  \label{eq:j-definition}
  j(\zeta)
  &=
  -4\sum_{n}q_n\delta(\zeta-q_n)
  \\
  &=
  4\sum_{n}\tau_n\sqrt{1-\tau_n} \delta(\tau_n - \frac{4\zeta}{(1+\zeta)^2})
  \,.
\end{align}
The result is particularly simple for diffusive junctions,
$j(\zeta)=\theta(1+\zeta)$, as the transmission distribution
$\propto1/(\tau\sqrt{1-\tau})$ cancels the prefactor. Similar result
applies also to the analogously defined $\tilde{j}$ corresponding to
$\tilde{f}$. For diffusive junctions,
$\tilde{j}(\zeta)=\theta(\zeta)$.

From the above discussion, and the symmetries
$(\hat{y}^{R/A})^\dagger=\hat{\tau}_3(\hat{y}^{A/R})^{-1}\hat{\tau}_3$,
$(\check{y}^K)^\dagger=-\hat{\tau}_3(\check{y}^{-1})^K\hat{\tau}_3$,
it follows that
\begin{align}
  \label{eq:Q-rewrite}
  \dot{Q}_{R,dc}
  &=
  \int_{-\infty}^\infty
  \dd{E}
  E
  \sum_n
  q_n
  \Re\tr\check{\sigma}_1\Bigl(\frac{1}{q_n + \check{y}}\Bigr)_{00}
  \\\notag
  &=
  \int_{-\infty}^\infty
  \dd{E}
  E
  \sum_n
  q_n
  \Re\tr\Bigl(\frac{1}{q_n + \hat{y}^R}\hat{y}^K\frac{1}{q_n + \hat{y}^A}\Bigr)_{00}
  \,,
\end{align}
which is a form useful for analytic considerations.  In particular,
this can be rewritten in the form used in Sec.~\ref{sec:qp-transport},
with
\begin{align}
  \label{eq:y-functions}
  \frac{Y_{RR,k}}{R}
  &=
  -\Re\tr
  \sum_n
  \Bigl(\frac{q_n}{q_n + \hat{g}^R_L\hat{g}^R_R}\hat{g}_{L}^R\Bigr)_{0,k}
  \\\notag
  &\qquad\times
  \Bigl((\hat{g}_{R,0}^R-\hat{g}_{R,0}^A)\frac{1}{q_n + \hat{g}^A_L\hat{g}^A_R}\Bigr)_{k,0}
  \,,
  \\
  \frac{Y_{RL,k}}{R}
  &=
  \Re\tr
  \sum_n
  \Bigl(\frac{q_n}{q_n + \hat{g}^R_L\hat{g}^R_R}e^{i\phi\tau_3}(\hat{g}_{L,0}^R-\hat{g}_{L,0}^A)\Bigr)_{0,k}
  \\\notag
  &\qquad\times
  \Bigl(e^{-i\phi\tau_3}\hat{g}_{R,0}^A\frac{1}{q_n + \hat{g}^A_L\hat{g}^A_R}\Bigr)_{k,0}
  \,.
\end{align}
The results corresponding to $\dot{Q}_{L,dc}$ are found by exchanging
$\phi\leftrightarrow-\phi$ and $L\leftrightarrow{}R$.  A sum rule
applies,
\begin{gather}
  \sum_k\frac{Y_{RR,k} + Y_{RL,k}}{R}
  =
  \\\notag
  \Re\tr
  \sum_n
  \Bigl(\frac{q_n}{q_n + y^R}[y^R - y^A]\frac{1}{q_n + y^A}\Bigr)_{00}
  =
  0
  \,,
\end{gather}
where the second step follows from
$(\hat{y}^{A})^\dagger=(\hat{g}_R^{A})^\dagger(\hat{g}_L^{A})^\dagger=\hat{\tau_3}\hat{g}_{R}^{R}\hat{y}^R\hat{g}_{R}^{R}\hat{\tau_3}$,
and the fact that $\hat{g}_R$ is diagonal in the energy space.

Finally, let us point out how the heat current in diffusive junctions
is phase-independent in the absence of bias. In the absence of
spectral broadening, quasiparticle heat current can only originate
from region $|E|>|\Delta_L|,|\Delta_R|$. From the form of the
equilibrium Green functions~\eqref{eq:g-equilibrium} it follows that
for $|E|>|\Delta_L|,|\Delta_R|$,
\begin{align}
  \label{eq:yeq}
  \hat{g}^A_{L/R}(E) &= -\hat{g}^R_{L/R,0}(E)
  \,,
  \\
  \check{y}_{\rm eq}(E)
  &=
  \begin{pmatrix}
    \hat{y}_{\rm eq}^R(E) & 2\hat{y}_{\rm eq}^R(E)[h_R-h_L]
    \\
    0 & \hat{y}_{\rm eq}^R(E)
  \end{pmatrix}
  \,.
\end{align}
Therefore, for this energy range,
\begin{align}
  \tr \sigma_1\frac{1}{\zeta+\check{y}_{\rm eq}(E)}
  =
  \tr \frac{2\hat{y}_{\rm eq}^R(E)}{(\zeta+\hat{y}_{\rm eq}^R(E))^2} (h_R - h_L)
  \,.
\end{align}
Integrating over the $\tilde{j}$ distribution yields
\begin{align}
  \label{eq:diffusive-coincidence}
  \int_{0}^{\infty}\dd{\zeta}\frac{2\hat{y}_{\rm eq}^R(E)}{(\zeta+\hat{y}_{\rm eq}^R(E))^2}
  =
  2
  \,,
\end{align}
which indeed has no phase modulation.  It is also clear that phase
modulation in general prevails if the distribution $\tilde{j}(\zeta)$
is nontrivial.

\section{DC voltage bias recurrence}

\label{sec:dc-recurrence}

We proceed now to evaluate the cumulant generating function
of energy exchange \cite{snyman2008-kam,kindermann2004-sht,virtanen2014-fhc}
\begin{align}
  {\cal S}_R(u) &= 
  \frac{1}{2}
  \sum_n\Tr\ln[1 + \frac{\tau_n}{4}([\check{g}_L,\check{g}_R(u)]_+ - 2)]
  \\
  &=
  \sum_n \Tr\ln[q_ne^{i\phi\hat{\tau}_3/2}\check{g}_{L,0}e^{-i\phi\hat{\tau}_3/2} 
  \\\notag
  &\qquad+ e^{-i\phi\hat{\tau}_3/2}e^{iu\epsilon\sigma_1/2}\check{g}_{R,0}e^{-iu\epsilon\sigma_1/2}e^{i\phi\hat{\tau}_3/2}]
\end{align}
in the dc voltage-biased case. One can simplify the problem by reusing
the approach of Ref.~\onlinecite{cuevas2004-dti}, where a similar
generating function for the charge rather than the heat current is
discussed.  The key idea is to make use of the fact that the sum of
the Green functions is a block-tridiagonal matrix in energy space, so
that one can apply a determinant recursion formula
\begin{align}
  \det X
  &=
  \prod_{j=1}^N \det a_j
  \,,
  \\
  a_1 &= X_{1,1}
  \,,
  \\
  \label{eq:a-recurrence}
  a_{j+1} &= X_{j+1,j+1} - X_{j+1,j} a_{j}^{-1} X_{j,j+1}
  \,.
\end{align}
In the case here, it is useful to work in a gauge where the
superconducting terminals are biased to $\pm{}V/2$, so that the
frequency components read
\begin{align}
  [\check{g}_L]_{ij}
  &=
  \begin{pmatrix}
    \delta_{ij}\breve{g}_{L}(E_i) & \delta_{j,i+1}\breve{f}_{L}(E_i) \\
    -\delta_{i,j+1}\breve{f}_{L}(E_i) & -\delta_{ij}\breve{g}_{L}(E_{i}-V)
  \end{pmatrix}
  \,,
  \\
  [\check{g}_R]_{ij}
  &=
  \begin{pmatrix}
    \delta_{ij}\breve{g}_{R}(E_{i}-V) & \delta_{i,j+1}\breve{f}_{R}(E_i) \\
    -\delta_{j,i+1}\breve{f}_{R}(E_i) & -\delta_{ij}\breve{g}_{R}(E_i)
  \end{pmatrix}
  \,,
\end{align}
where $E_n = E_0 + nV$, and we work in Nambu$\otimes$Keldysh instead
of Keldysh$\otimes$Nambu:
\begin{align}
  \breve{g}_{L}
  &=
  \begin{pmatrix}
    \cosh\theta_{L} & 2\Re[\cosh\theta_{L}]\tanh\frac{E}{2T_L} \\
    0 & -\cosh\theta_{L}^*
  \end{pmatrix}
  \,,
  \\
  \label{eq:gr-unitary}
  \breve{g}_{R}
  &=
  e^{iEu\breve{\sigma}_1/2}
  \begin{pmatrix}
    \cosh\theta_{R} & 2\Re[\cosh\theta_{R}]\tanh\frac{E}{2T_R} \\
    0 & -\cosh\theta_{R}^*
  \end{pmatrix}
  \\\notag
  &\qquad\times
  e^{-iEu\breve{\sigma}_1/2}
  \,,
\end{align}
where
$\theta_{L/R}=\arctanh[\Delta_{L/R}/(E+i0^+)]$. Expressions
for $\breve{f}$ are obtained by replacing $\cosh\mapsto\sinh$.

In the recursion formula for the generating function, we therefore
have
\begin{align}
  X_{j,j} 
  &= 
  \begin{pmatrix}
    q_n \breve{g}_{L}(E_j) + \breve{g}_{R}(E_{j-1}) & 0 \\
    0 & -q_n \breve{g}_{L}(E_{j-1}) - \breve{g}_{R}(E_j)
  \end{pmatrix}
  \\
  X_{j+1,j} 
  &= 
  \begin{pmatrix}
    0 & \breve{f}_{R}(E_j) \\
    -q_n\breve{f}_{L}(E_j) & 0
  \end{pmatrix}
  \\
  X_{j,j+1} 
  &= 
  \begin{pmatrix}
    0 & q_n \breve{f}_{L}(E_j) \\
    -\breve{f}_{R}(E_j) & 0
  \end{pmatrix}
  \,.
\end{align}
Clearly, $a_j$ stays diagonal in the recursion.

Finding the generating function now reduces to solving
a pair of $2\times2$ matrix recurrences
\begin{subequations}
\label{eq:v-bias-recursion}
\begin{align}
  \breve{v}(E+V,u)
  &=
  q_n\breve{g}_{L}(E+V) + \breve{g}_{R}(E,u)
  \\\notag&\qquad
  + \breve{f}_{R}(E,u) [\breve{w}(E,u)]^{-1} \breve{f}_{R}(E,u)
  \,,
  \\
  \breve{w}(E+V,u)
  &=
  -q_n\breve{g}_{L}(E) - \breve{g}_{R}(E+V,u)
  \\\notag&\qquad
  + q_n^2\breve{f}_{L}(E) [\breve{v}(E,u)]^{-1} \breve{f}_{L}(E)
  \,,
\end{align}
\end{subequations}
and
\begin{align}
  {\cal S}_R(u)
  &=
  t_0
  \sum_{q_n}
  \int_{-\infty}^\infty\frac{\dd{E}}{2\pi}
  \ln\frac{\det \breve{v}(E,u) \det \breve{w}(E,u)}{(1+q_n)^4}
  +
  C
  \,,
\end{align}
where a normalization was added to make the result convergent;
this does not affect observables.
Differentiation yields directly the average heat current,
$\partial_u{\cal S}_R\rvert_{u=0}=i\dot{Q}_{R,dc}$.

The recursion can be solved numerically in a straightforward way, as
$\breve{f}\to0$ for $|E|\to\infty$.  Analytical results can be
obtained in some particular limiting cases.

\subsection{Tunnel limit}

We can consider the solution of the recursion in the tunnel limit
$q\to\infty$.  There, to order ${\cal O}(q^0)$, it reads
\begin{align}
  \breve{v}(E+V)
  &\simeq
  q_n \breve{g}_L(E+V) + \breve{g}_R(E)
  \,,
  \\
  \breve{w}(E+V)
  &
  \simeq
  -q_n \breve{g}_L(E) - \breve{g}_{R}(E+V)
  \\\notag&\quad
  + q_n \breve{f}_L(E)\breve{g}_L(E)^{-1}\breve{f}_L(E)
  \\\notag&\quad
  - \breve{f}_L(E)\breve{g}_L(E)^{-1}\breve{g}_R(E-V)\breve{g}_L(E)^{-1}\breve{f}_L(E)
  \\\notag
  &=
  -q_n \breve{g}_L(E)^{-1} - \breve{g}_R(E+V) 
  \\\notag&\quad
  - \breve{f}_L(E)\breve{g}_L(E)^{-1}\breve{g}_R(E-V)\breve{g}_L(E)^{-1}\breve{f}_L(E)
  \,,
\end{align}
where we used the normalization conditions $\breve{g}^2 - \breve{f}^2 = \breve{1}$,
$\breve{g}\breve{f}=\breve{f}\breve{g}$.
From this,
\begin{gather}
  \ln\det\breve{v}(E,u)\breve{w}(E+V,u)
  \\\notag
  \simeq
  \frac{1}{q_n}\Tr
  \breve{g}_L(E)[\breve{g}_R(E+V,u)+\breve{g}_R(E-V,u)]
  +
  C
  \,.
\end{gather}
This yields the result for heat transport by tunneling quasiparticles,
also obtained in Ref.~\onlinecite{golubev2013-htt},
\begin{align}
  {\cal S}_R(u)
  &\simeq
  \frac{t_0}{\pi}
  \sum_{n}\tau_n
  \int_{-\infty}^\infty\dd{E}
  N_L(E)N_R(E - V)
  \\\notag
  &\;\times
  \Bigl\{
  [e^{i(E - V)u}-1]F_R(E - V)[1-F_L(E)]
  \\\notag
  &\qquad
  +
  [e^{-i(E -V)u}-1]F_L(E)[1-F_R(E - V)]
  \Bigr\}
  \,,
\end{align}
where $N=\Re\cosh\theta$ are the densities of states on both sides,
and $F_{L/R}$ Fermi functions.

\subsection{Zero bias}

In the zero-bias limit $V\to0$, the results coincide with averaging
the corresponding d.c. formulas in
Refs.~\onlinecite{zhao2003-phase,virtanen2014-fhc} over the phase
difference $\varphi=0\ldots2\pi$.  However, in this case it is also
possible to solve the recurrence relation exactly.

We first observe that a factorization applies:
\begin{align}
  \breve{g}_L(E) &= g_L(E) \breve{h}_L(E)
  \,,
  \quad
  g_L(E)
  =
  \cosh\theta_L(E)
  \,,
  \\
  \breve{h}_L(E) &= \begin{cases}
    1 
    \,, 
    &
    |E|<|\Delta_L|
    \,,
    \\
    \begin{pmatrix}
      1 & 2 h_L \\ 0 & -1
    \end{pmatrix}
    &
    |E|>|\Delta_L|
  \end{cases}
  \,,
\end{align}
and similarly for $\breve{f}$. For $L\mapsto{}R$, the $\breve{h}_R$ matrix
obtains the additional $u$-dependent unitary transformation \eqref{eq:gr-unitary}.

Importantly, $\breve{h}_R^2 = 1$ and $\breve{h}_L^2 = 1$.
Defining $\breve{x}=\breve{v}\breve{h}_L$ and 
$\breve{y}=\breve{w}\breve{h}_R$, we have
\begin{align}
  \breve{x}
  &=
  q_n g_L + g_R \breve{h}_R\breve{h}_L + f_R^2\breve{y}^{-1}\breve{h}_R\breve{h}_L
  \,,
  \\
  \breve{y}
  &=
  -q_n g_L [\breve{h}_R\breve{h}_L]^{-1}  - g_R + q_n^2 f_L^2 \breve{x}^{-1}[\breve{h}_R\breve{h}_L]^{-1}
  \,.
\end{align}
The equations projected to the eigenspaces of $\breve{h}_R\breve{h}_L$
have no matrix structure, and can be solved. The eigenvalues $\mu$ of
$\breve{h}_R\breve{h}_L$ satisfy
\begin{align}
  \mu^{-1}+\mu-2
  &=
  4(e^{iuE}-1)F_L(1-F_R) 
  \\\notag
  &\qquad
  + 4(e^{-iuE}-1)F_R(1-F_L)
  \,.
\end{align}
As a consequence, the generating function can be found in closed form,
although the general expressions are not particularly illuminating.

For simplicity, let us consider $\tau=1$ and $\Delta_L=\Delta_R$.
In this case,
\begin{align}
  \label{eq:tau1-statistics}
  \ln\det[\breve{v}\breve{w}]
  &=
  2
  \ln\Bigr[
  \frac{\cosh^{-4}\frac{\theta}{2}}{8}
  \bigl(
    \mu^{-1} + \mu + 1 + \cosh(2\theta)
    \\\notag
    & + (1+\mu)\mu^{-1/2}\sqrt{\mu^{-1}+\mu+2\cosh(2\theta)}
  \bigr)
  \Bigl]
  +
  C
  \,.
\end{align}
This $V\to0$ heat transport statistics deviates from the stationary
Levitov--Lesovik form \cite{levitov1993-cdi,levitov96} due to the
averaging over the slowly varying phase $\varphi(t)=2Vt$ in the long
time limit.

Expanding in $u$, we find
\begin{align}
  \ln\det[\breve{v}\breve{w}]
  =
  C
  -
  2 i u E  (f_L - f_R) \frac{1}{\cosh\theta}
  +
  {\cal O}(u^{3/2})
  \,,
\end{align}
so that the heat current is
\begin{align}
  \dot{Q}_{R,dc}
  =
  \frac{1}{\pi}\int_{-\infty}^\infty\dd{E}
  E
  \Re\frac{\sqrt{E^2 - \Delta^2}}{|E|}
  [f_R(E) - f_L(E)]
  \,.
\end{align}
We can verify that this coincides with the average of the d.c. result in
Ref.~\onlinecite{zhao2003-phase}:
\begin{align}
  Y
  =
  \int_0^{2\pi}\frac{\dd{\varphi}}{2\pi}
  \frac{
    (E^2 - \Delta^2)(E^2 - \Delta^2\cos^2\frac{\varphi}{2})
  }{
    [E^2 - \Delta^2(1-\sin^2\frac{\varphi}{2})]^2
  }
  =
  \frac{\sqrt{E^2 - \Delta^2}}{|E|}
  \,.
\end{align}
For comparison with the inset of Fig.~\ref{fig:qpc-Y-V-dep}, let us also state the result
for $\tau=1/2$:
\begin{align}
  \label{eq:tau05-zerobias}
  Y(E>\Delta)
  =
  \frac{
    2E^2\sqrt{4E^4 - 6\Delta^2E^2 + 2\Delta^4}
  }{
    (2E^2 - \Delta^2)^2
  }
  \,,
\end{align}
which coincides with numerics.

\section{Small ac phase bias}

\label{sec:ac-bias-expansion}

In the case of a small ac phase bias, $\phi=\varphi_0/2 + a$, we can
expand
\begin{align}
  e^{i\phi\tau_3}\approx e^{i\varphi_0\hat{\tau}_3/2}(1 + i a \hat{\tau}_3 - \frac{1}{2} a^2) + \ldots
\end{align}
The corresponding expansion for the expressions appearing in the heat current is
\begin{align}
  \label{eq:ac-expansion}
  \frac{1}{q_n + \check{y}}
  &\simeq
  \frac{1}{q_n + \check{y}_0}
  +
  \frac{1}{q_n + \check{y}_0}
  [ia\hat{\tau}_3, \check{g}_{L,0}]\check{g}_{R,0}
  \frac{1}{q_n + \check{y}_0}
  \\\notag
  &\qquad
  -
  \frac{1}{q_n + \check{y}_0}
  \check{X}
  \frac{1}{q_n + \check{y}_0}
  \\
  \check{X}
  &=
  -
  \frac{1}{2}
  a^2
  \check{g}_{L,0}\check{g}_{R,0}
  -
  \frac{1}{2}
  \check{g}_{L,0}a^2\check{g}_{R,0}
  +
  \hat{\tau}_3
  a\check{g}_{L,0}\hat{\tau}_3a\check{g}_{R,0}
  \\\notag
  &
  +
  [a\hat{\tau}_3, \check{g}_{L,0}]_-\check{g}_{R,0}
  \frac{1}{q_n + \check{y}_0}
  [a\hat{\tau}_3, \check{g}_{L,0}]_-\check{g}_{R,0}
  \,,
\end{align}
where $\hat{g}_{L,0}$ contains the phase $\varphi_0$ and
$\check{y}_0=\hat{g}_{L,0}\hat{g}_{R,0}$.
For harmonic drive, $a(t)=2s\cos(\omega_0t)$, we can evaluate the
leading term in the expansion in $s$ in closed form.

It is useful to identify the stationary Andreev bound states (ABS):
\begin{gather}
  \frac{
    1
  }{
    q + \check{y}_0^{R/A}(E)
  }
  =
  \frac{
    \hat{B}_+
  }{
    E \pm i\eta - E_A
  }
  +
  \frac{
    \hat{B}_-
  }{
    E \pm i\eta + E_A
  }
  +
  \frac{1}{1 + q}
  \,,
\end{gather}
where $E_A=|\Delta|\sqrt{1-\tau\sin^2(\varphi_0/2)}$ are the ABS
energies, and $\hat{B}_\pm$ 2x2 matrixes independent of $E$.  The
contributions to the heat current separate to $\delta$-contributions
from $E\pm\omega=\pm{}E_A$ corresponding to processes
involving the bound states, and principal value parts corresponding to
continuum processes.

Straightforward calculation for the principal value part yields,
\begin{align}
  \label{eq:general-ac-analytic-1}
  Y_{RL,0} &= \delta Y_{RL,0} + \frac{1}{r}\sum_n q_n D_0(q_n)
  \\
  \delta Y_{RL,0} &= -s^2\frac{1}{r}\sum_\pm \sum_n q_n D^{\pm}_1(q_n) \theta_0
  \\
  Y_{RL,\pm} &= s^2\frac{1}{r}\sum_n q_n [D^{\pm}_2(q_n) + D^{\pm}_3(q_n)]\theta_0\theta_{\pm\omega}
  \\
  Y_{RR,\pm} &= s^2\frac{1}{r}\sum_n q_n [D^{\pm}_2(q_n) - D^{\pm}_3(q_n)]\theta_0\theta_{\pm\omega}
\end{align}
where $r=\sum_n\tau_n$, $\theta_\omega=\theta((E+\omega)^2 - \Delta^2)$, and 
\begin{align}
  D_0(q)
  &=
  2\tr\frac{\hat{y}_0}{(q + \hat{y}_0)^2}
  \,,
  \\
  D_1^\pm(q)
  &=
  2\tr\biggl(
  \hat{g}_{R,0} \frac{4q\hat{y}_0}{(q+\hat{y}_0)^3}\hat{\tau}_3\frac{1}{q+\hat{y}_{\pm\omega}}\hat{g}_{L,\pm\omega}\hat{\tau}_3
  \\\notag
  &\qquad
  +\frac{(q-\hat{y}_0)\hat{y}_0}{(q+\hat{y}_0)^3}
  \hat{\tau}_3\frac{q - \hat{y}_{\pm\omega}}{q + \hat{y}_{\pm\omega}}
  \hat{\tau}_3
  \biggr)
  \,,
  \\
  D_2^\pm(q)
  &=
  \tr\biggl(
  \hat{g}_{R,0} 
  \frac{1}{q + \hat{y}_0}
  \hat{\tau}_3
  \frac{1}{q + \hat{y}_{\pm\omega}}
  \hat{g}_{L,\pm\omega}
  \hat{\tau}_3
  \biggl)
  \,,
\end{align}
\begin{align}
  \label{eq:general-ac-analytic-n}
  D_3^\pm(q)
  &=
  \tr\biggl(
  \hat{g}_{R,0}
  \frac{q - \hat{y}_0}{(q + \hat{y}_0)^2}
  \hat{\tau}_3
  \frac{q - \hat{y}_{\pm\omega}}{(q + \hat{y}_{\pm\omega})^2}
  \hat{g}_{L,\pm\omega}
  \hat{\tau}_3
  \\\notag
  &\qquad
  +
  \frac{
    4q\hat{y}_0
  }{
    (q + \hat{y}_0)^2
  }
  \hat{\tau}_3
  \frac{
    \hat{y}_{\pm\omega}
  }{
    (q + \hat{y}_{\pm\omega})^2
  }
  \hat{\tau}_3
  \biggr)
  \,.
\end{align}
We have here used additional symmetries arising from assuming
$\Delta_L=\Delta_R$.  Above,
\begin{align}
  \hat{g}_{R,\omega}
  &=
  \frac{
    1
  }{
    \sqrt{(E+\omega)^2 - \Delta^2}
  }
  \begin{pmatrix}
    |E+\omega| & \Delta \\ -\Delta & -|E+\omega|
  \end{pmatrix}
  \,,
  \\
  \hat{g}_{L,\omega}
  &=
  \frac{
    1
  }{
    \sqrt{(E+\omega)^2 - \Delta^2}
  }
  \begin{pmatrix}
    |E+\omega| & e^{i\varphi_0}\Delta \\ 
    -e^{-i\varphi_0}\Delta & -|E+\omega|
  \end{pmatrix}
  \\
  \hat{y}_\omega
  &=
  \hat{g}_{L,\omega} \hat{g}_{R,\omega}
  \,.
\end{align}
Evaluating the result for $\tau=1$ ($q=1$) yields
Eqs.~\eqref{eq:ac-phase-y-1},\eqref{eq:ac-phase-y-2}.  Explicit
expressions for general $\tau$ can be written, but are too lengthy to
be useful.  

We can also observe that the coincidence
\eqref{eq:diffusive-coincidence} that made phase oscillations
disappear in the diffusive limit for the stationary heat current does
not occur for the time-dependent modifications above. While $D_0(q)$
averages to a phase independent constant when integrating over the $q$
distribution, the other coefficients retain their phase dependence.

In the situation $T_L=T_R=T$, $\Delta_L=\Delta_R=\Delta$ only
$D_2^\pm$ contributes to the principal value part. The reason why the
result in this case is simpler, is that heat currents are then
directly connected to the total dissipated power, $P=\overline{I(t)
  V(t)}$, via $P=\dot{Q}_L+\dot{Q}_R=2\dot{Q}_R$.  The power can be
obtained by calculating $I(t)$ to \emph{first} order in $a$, i.e.,
from the frequency-dependent admittance of the junction.  This is done
via direct linear response approach in Ref.~\onlinecite{kos2013-fda},
and the results coincide with the above formulation.

\end{document}